\documentclass[prb,preprint,groupedaddress,showpacs,showemail]{revtex4}

\usepackage{graphicx}

\begin{document}

\title{Role of dipolar and exchange interactions in the positions
and widths of EPR transitions for the 
single-molecule magnets Fe$_{\mathbf 8}$ and Mn$_{\mathbf 12}$}
\author{Kyungwha Park$^{1,2,}$}\email{park@csit.fsu.edu} 
\author{M.~A.\ Novotny$^{3,}$}\email{man40@ra.msstate.edu} 
\author{N.~S.\ Dalal$^{2,}$}\email{dalal@chem.fsu.edu}
\author{S.\ Hill$^{4,}$}\email{hill@phys.ufl.edu}
\author{P.~A.\ Rikvold$^{1,5,}$}\email{rikvold@csit.fsu.edu}
\affiliation{
$^1$School of Computational Science and 
Information Technology, Florida State University, Tallahassee, 
Florida 32306 \\
$^2$Department of Chemistry and Biochemistry, 
Florida State University,
Tallahassee, Florida 32306 \\
$^3$Department of Physics and Astronomy, Mississippi 
State University, Mississippi State, Mississippi 39762 \\
$^4$Department of Physics, 
University of Florida, Gainesville, Florida 32611 \\
$^5$Center for Materials Research and 
Technology and Department of Physics, 
Florida State University, Tallahassee, Florida 32306
}
\date{\today}


\begin{abstract}  

We examine quantitatively the temperature dependence of the linewidths and
line shifts in electron paramagnetic resonance experiments
on single crystals of the single-molecule magnets Fe$_8$ and Mn$_{12}$,
at fixed frequency, with an applied magnetic field along the easy axis. 
We include inter-molecular spin-spin interactions (dipolar and exchange) 
and distributions in both the uniaxial
anisotropy parameter $D$ and the Land\'{e} $g$-factor. 
The temperature dependence of the linewidths and the line shifts 
are mainly caused by the spin-spin interactions.  
For Fe$_8$ and Mn$_{12}$, the temperature dependence of the 
calculated line shifts and linewidths agrees well with the trends of
the experimental data. 
The linewidths for Fe$_8$ reveal a stronger 
temperature dependence than those for Mn$_{12}$,
because for Mn$_{12}$ a much wider distribution in $D$ overshadows
the temperature dependence of the spin-spin interactions.
For Fe$_8$, the line-shift analysis suggests two competing interactions:
a weak ferromagnetic exchange coupling between neighboring molecules and
a longer-ranged dipolar interaction. 
This result could have implications for ordering in Fe$_8$
at low temperatures.
\end{abstract}
\maketitle


\section{Introduction}

Single-molecule magnets (SMM's) consist of identical molecules, 
each of which made up of several magnetic ions surrounded 
by many different species of atoms. 
A single molecule of the SMM's
Mn$_{12}$-acetate\cite{LIS80} and Fe$_8$\cite{WIEG84} has 
an effective ground-state spin of $S=10$ and a strong crystal-field 
anisotropy. A zero-field energy barrier against magnetization reversal 
is approximately 65 K (30 K) for uniaxial Mn$_{12}$ (biaxial Fe$_8$).
\cite{BARR97,HILL98,PERE98,MIRE99,CACI98}
Despite their large effective spin, these single-molecule magnets 
have shown quantum coherence\cite{GARG93,GUNT95} and 
quantum tunneling between the energy levels of the 
two potential wells.\cite{GUNT95,VILL94,CHUD98} 
Although dipolar interactions between different molecules 
are weak, in the low-temperature limit and near zero applied field, 
a dipolar interaction could stimulate the quantum tunneling and
thus explain the non-exponential magnetization relaxation observed
at early times.\cite{PROK98,OHM98,CUCC99,WERN99,MIYA01,LIU01}

Recently, electron paramagnetic resonance (EPR) experiments
on single crystals of Fe$_8$ and Mn$_{12}$ have revealed
interesting effects in the widths and positions of the EPR peaks 
as functions of energy level, resonance frequency, and temperature
when the applied field is along the easy axis.
\cite{HILL98,BLIN01,MACC01,PARK02,HILL02} 
For both Fe$_8$ and Mn$_{12}$, at fixed frequency, the linewidths 
increase with decreasing energy levels (the largest linewidth corresponding
to the transition between the ground state and the first excited state),
and for a particular transition the linewidths increase with decreasing 
frequency. On the other hand, the details of the temperature dependence 
of the linewidths for the two materials are quite different.
For Mn$_{12}$, the linewidths increase smoothly with increasing temperature,
showing a rather weak temperature dependence.
For Fe$_8$, for transitions at low resonant fields, 
the linewidths increase sharply with temperature 
at low temperatures, reach a maximum, and then decrease slowly with 
temperature at higher temperatures. The exception is
the transition associated with the ground state, for which
the linewidth decreases with increasing temperature
in the whole temperature range studied (2$-$50 K).
On the other hand, for the transitions with high resonant fields,
the linewidths increase monotonically with increasing temperature.
For Fe$_8$, the line positions change non-monotonically with
energy level and temperature.

In Ref.\onlinecite{PARK02}, which was our first attempt to understand
the energy level and resonance frequency dependence of the linewidths, we 
showed that for both single crystals of Fe$_8$ and Mn$_{12}$ the distribution
in the uniaxial anisotropy parameter $D$ of the single-spin Hamiltonian,
caused by defects in the samples, contributes substantially to 
the inhomogeneous linewidths at constant temperature. 
This was also recently supported 
by terahertz spectroscopy for Mn$_{12}$.\cite{PARKS01}
The microscopic origin of the distribution in $D$ has not yet been 
fully understood.\cite{CHUD01,CORN01} The analysis further
showed that for Fe$_8$ the dipolar interactions between molecules
contribute to the linewidths as significantly as the distribution in $D$,
while for Mn$_{12}$ their contribution is less significant.\cite{PARK02}
In a recent millimeter-wave study on Fe$_8$,\cite{MUKH01} 
the dipolar field may have been overestimated because the distribution in $D$ 
was not included. 

Since the approximations made in Ref.\onlinecite{PARK02} reasonably 
well explained the linewidth behavior at fixed
temperature, in the present paper we investigate the temperature 
dependence of the linewidths and line positions.
To explain this additional feature, we take into account the 
inter-molecular spin-spin interactions (exchange and dipolar),
as well as distributions in the uniaxial anisotropy parameter $D$ and 
the Land\'{e} $g$-factor. We find that the distributions in $D$ and $g$ 
do not contribute to shifts in line positions with temperature, but 
the exchange and dipolar interactions produce nonzero local fields 
which are temperature dependent (at low temperatures), so that they can 
make the line positions change with temperature. Thus, to explain the 
{\it temperature} dependence of the measured line shifts, both the 
{\it exchange} and dipolar interactions are needed. Without the exchange 
interaction, we cannot explain the observed non-monotonic 
line shifts with temperature. 
This was not included in our earlier study, Ref.\onlinecite{PARK02}.
Although the exchange interaction was not considered in our earlier
study, its effect on the linewidths is 
negligible as long as the magnitude of the exchange constant is much 
smaller than the linewidths.\cite{AABB} 

This paper is organized as follows. A brief summary of the experimental
procedures is presented in Sec.~II. The models for Fe$_8$ and
Mn$_{12}$ are described separately in Sec.~III. Sec.~IV describes
our calculated linewidths and line shifts vs temperature, and their
discussed in comparison with the experimental data. 
Our conclusions are provided in Sec.~V.

\section{Summary of Experimental Procedures}

All of the EPR experiments were performed on single crystals with the magnetic 
field aligned along or close to the direction of the easy magnetization axis,
in a temperature range from 2 K to 50 K. 
For Mn$_{12}$, this direction coincides with that of the longest 
dimension of the essentially needle-shaped crystals. For Fe$_8$, 
the direction was determined by measurement of the largest spread of 
the resonant field, by rotating the crystal around the approximately 
known orientation of the easy axis. The single crystals were prepared 
as described earlier.\cite{HILL98,PERE98,ACHE01-1,ACHE01-2,ACHE02}
EPR measurements were made in the 100$-$190 GHz range with a 
resonant microwave cavity system described by Hill et al.\cite{HILL99}, 
which enables observation of distortion-free EPR line shapes. The linewidths 
were determined by computer-fitting of the observed experimental spectra to 
either a Gaussian or a Lorentzian function for the spectra obtained at 
various temperatures. All the spectra were obtained by keeping the 
frequency fixed and sweeping the field to obtain the resonance peaks, as 
is usual in EPR spectroscopy.\cite{DALA82}

\section{Model}

For the examined single-crystal sample of Fe$_8$,\cite{HILL02} 
there are two sources of the EPR line {\it shifts}: temperature 
dependence of $D$ and the electronic spin-spin interactions
(dipolar and exchange interactions) between different molecules. 
There are also two sources of the EPR line {\it broadening}: the 
$D$-strain effect (distribution in $D$) and 
the spin-spin interactions.\cite{PARK02} 
For the examined Mn$_{12}$ sample,\cite{HILL02} the sources of
the line broadening are the $D$-strain, the $g$-strain,\cite{PILB}
and the spin-spin interactions.\cite{PARK02} 

In our model, hyperfine interactions are not considered 
for the following reasons: (i) The observed line shifts and
changes in the linewidths were much larger for Fe$_8$ than for Mn$_{12}$.
(ii) On the other hand, hyperfine fields are orders of magnitude
larger for Mn$_{12}$ (in which all nuclei have spins of $I=5/2$)
than for Fe$_8$ (98\% of Fe nuclei with $I=0$). (iii) Any residual
effect of the hyperfine fields was included in background linewidths.

Possible mis-alignment of the external field and a spread
in the in-plane fields are also not considered since
mis-alignment of the external field cannot provide
the temperature dependence of the linewidths and shifts.
We checked that for Fe$_8$ the effect of a distribution in the transverse 
anisotropy parameter $E$ on the linewidths was negligible.
We also found that a spread in the in-plane fields (possibly 
caused by nuclear spins) can give rise to
a distribution in the easy axis of each molecule, leading
to asymmetries in EPR lineshapes.\cite{PARK02-2,PARK02-3}

The model and technique used in this study are similar to
those in Ref.\onlinecite{PARK02}, except that we here take into
account the non-spherical sample shapes and the actual crystal structures. 
Therefore, we here summarize them only briefly, focusing on what causes 
the temperature dependence of the line shifts and linewidths 
for each source. The temperature dependence of $D$ is discussed
in Sec.~IV A.1. In our convention, the energy level $M_s=+10$ is 
the ground state when the field is applied along the positive $z$ axis, 
while in Refs.\onlinecite{MACC01,HILL02} the ground state is $M_s=-10$.
For clarity, we discuss Fe$_8$ and Mn$_{12}$ separately.

\subsection{Fe$_{\mathbf 8}$}

We consider an effective single-spin Hamiltonian 
which satisfies approximate $D_2$ symmetry,
\begin{eqnarray}
{\cal H}_{0} &=& -D S_z^2 - E(S_x^2 - S_y^2) 
- g \mu_B H_z S_z ~, \label{eq:ham}
\end{eqnarray}
where the uniaxial anisotropy parameter $D=0.288k_B$, 
the transverse anisotropy parameter $E=0.043k_B$,\cite{MACC01}
$g$ is the Land\'{e} $g$-factor which is close to 2, 
and $\mu_B$ is the Bohr magneton. Here $S_{\alpha}$ is the $\alpha$th 
component of the spin angular momentum operator, 
and $H_z$ is the longitudinal static applied magnetic field.
We assume that the longitudinal magnetic field is 
applied along the easy axis (the $z$ axis), and
we ignore the small transverse anisotropy terms ($E$ terms) in
calculating the linewidths. Thus, the energy level $M_s$ is a
good quantum number of the spin operator $S_z$.
According to our convention, the ground state is $M_s=+10$.

Using the density-matrix equation \cite{BLUM} 
with the Hamiltonian, Eq.~(\ref{eq:ham}),
and an interaction between the spin system and
an oscillating transverse field, we calculate the power absorption
between the levels $M_s$ and $M_s-1$ for a fixed value of the uniaxial
anisotropy parameter $D$. In the power absorption, 
the line-shape function includes a natural linewidth, which is 
a function of temperature. Next we calculate the average power 
absorption with a Gaussian distribution in $D$, where $D$ 
itself is assumed to be temperature independent. 
As a consequence, the line broadening
due to the $D$-strain effect becomes weakly temperature dependent
because of the temperature dependence of the natural linewidths.
To calculate the natural linewidths, we use the strength of 
the coupling between the spin system and a surrounding phonon 
heat bath obtained in Ref.\onlinecite{LEUE00-2}. For example,
for $M_s=+10$, the order of magnitude of the natural 
linewidths is several to several tens of gauss at 
temperatures below several tens of kelvin, and the widths 
increase with decreasing $M_s$.

The spin-spin interactions are calculated separately with $D$ fixed,
and then combined with the $D$-strain effect to obtain the total
linewidths. At low temperatures, all energy levels are not equally populated,
and the populations of the levels change with temperature according to
the Boltzmann factor. Thus, the local field on a particular molecule
caused by surrounding molecules changes with temperature. 
Therefore, the spin-spin interactions are mainly responsible for 
the temperature dependence of both the line shifts and the linewidths.
\cite{AABB,MCMI60} To calculate the line shifts and the 
line broadening due to the spin-spin interactions, 
we use a multi-spin Hamiltonian which commutes with
$\sum_{j} S_{j}^z$, where the sum runs over all molecules.
Details of the technique can be found in Ref.\onlinecite{MCMI60}.
\begin{eqnarray}
{\cal H}^{\mathrm tot}&=&\sum_{i}[\: {\cal H}_{0i} + V_{i}(t) \:] \:
+ {\cal H}^{(1)} \: , \\
{\cal H}^{(1)} &\equiv& 
{\cal H}^{\mathrm dipole} + {\cal H}^{\mathrm exch} \:, \\
{\cal H}^{\mathrm dipole}&=&\frac{1}{2} \sum_{jk}\,^{\prime} A_{jk} 
( \vec{S}_j \cdot \vec{S}_{k} - 3  S_{j}^z S_{k}^z ) \:, \: \:
A_{jk} \equiv \left( \frac{\mu_0}{4 \pi} \right)
\frac{(g \mu_B)^2}{2 r_{jk}^3} (3 \zeta_{jk}^2 - 1) \:, 
\label{eq:dp} \\
{\cal H}^{\mathrm exch}&=& \: \frac{1}{2} \sum_{jk}\,^{\prime} 
J_{jk} \: \vec{S}_j \cdot \vec{S}_{k} \:,
\label{eq:ex}
\end{eqnarray}
where ${\cal H}_{0i}$ is the single-spin Hamiltonian for the $i$th 
molecule, the sum $\sum_{i}$ runs over all molecules, 
and $V_i(t)$ is the interaction 
between the $i$th molecule and the oscillating transverse field. Here 
${\cal H}^{\mathrm dipole}$ is the dipolar interaction between the molecules,
and $\zeta_{jk}$ are the direction cosines of the vector 
between molecules $j$ and $k$ ($\vec{r}_{jk}$) relative to the 
easy axis ($z$ axis). The sum $\sum_{jk}^{\prime}$ runs over
all molecules, so that any two indices are not the same.
${\cal H}^{\mathrm exch}$ is the isotropic 
exchange interaction between the spins of nearest-neighbor molecules
where the exchange coupling constant $J_{ij}$ is $J$
if the $i$th and $j$th spins are nearest neighbors and zero otherwise.
It is reasonable to assume that $\sum_{i} V_i(t)$ is much smaller
than the dipolar and exchange interactions, 
which again are much smaller than the
sum of the single-spin Hamiltonians. Since the field is swept at constant
frequency, the energy levels change with the sweeping field. 
We neglect slight changes of the energy levels during resonances
and use $H_{\mathrm res}=(h\nu - D(2M_s-1))/g\mu_B$,
where $\nu$ is an EPR frequency,
as the field $H$ in the spin Hamiltonian 
to calculate the energy levels for the particular resonance.

To calculate the ${\ell}$th moment of the resonant
field deviation, we formulate the ${\ell}$th moment 
for a frequency sweep and then convert it to a field sweep.
This is justifiable because we neglect the slight energy change
caused by the change of the field during a resonance.
The probability density function of the EPR frequency $\nu$ 
is given by
\begin{eqnarray}
F_{\nu}(\nu) &=& \frac{\sum_{n} \sum_{n^{\prime}}^{\Delta}
\{ \exp(-{{\cal E}_n}/k_B T) 
- \exp(-{{\cal E}_{n^{\prime}}}/k_B T) \} \:
|\langle n| \sum_{j} S_{j}^x | n^{\prime} \rangle|^2}
{\sum_{n} \sum_{n^{\prime}}^{+} \{ \exp(-{{\cal E}_n}/k_B T) 
- \exp(-{{\cal E}_{n^{\prime}}}/k_B T) \} \:
|\langle n| \sum_{j} S_{j}^x | n^{\prime} \rangle|^2} \:,
\label{eq:fhh}
\end{eqnarray}
where ${\cal E}_n$ is the energy eigenvalue of $\sum_j {\cal H}_{0j}
+{\cal H}^{(1)}$, 
$|n \rangle$ is the corresponding eigenvector, 
$\sum_{n^{\prime}}^{+}$ denotes the sum over all states 
$|n^{\prime} \rangle$ such that ${\cal E}_{n^{\prime}} \geq {\cal E}_n$,
and $\sum_{n} \sum_{n^{\prime}}^{\Delta}$ denotes the sum over all states
for which $h \nu \leq {\cal E}_{n^{\prime}}-{\cal E}_n 
\leq h (\nu + d\nu)$.
Using Eq.~(\ref{eq:fhh}), we calculate the ${\ell}$th moment, 
\begin{eqnarray}
\langle \nu^{\ell} \rangle &=& \frac{\sum_{n} \sum_{n^{\prime}}^{+}
({\cal E}_{n^{\prime}} - {\cal E}_n)^{\ell} \{ \exp(-{{\cal E}_n}/k_B T) 
- \exp(-{{\cal E}_{n^{\prime}}}/k_B T) \} \:
|\langle n| \sum_{j} S^x_{j} | n^{\prime} \rangle|^2}
{\sum_{n} \sum_{n^{\prime}}^{+} \{ \exp(-{{\cal E}_n}/k_B T) 
- \exp(-{{\cal E}_{n^{\prime}}}/k_B T) \} \:
|\langle n| \sum_{j} S^x_{j} | n^{\prime} \rangle|^2} \:, {\ell}=1,2,...,
\label{eq:lth_moment}
\end{eqnarray}
where the temperature dependence of the linewidths is included 
through the Boltzmann factors, and the eigenvalues of ${\cal H}^{(1)}$
can contribute to $({\cal E}_{n^{\prime}} - {\cal E}_n)^{\ell}$
and/or the Boltzmann factors. Assuming that the dipolar and exchange
interactions, ${\cal H}^{(1)}$, are much smaller than the thermal energy, 
we expand the Boltzmann factor into 
\begin{eqnarray}
\exp \left[-\frac{\sum_j {\cal H}_{0j} +{\cal H}^{(1)}}{k_B T} 
\right]&=&
\exp \left( -\frac{\sum_j {\cal H}_{0j}}{k_B T} \right) 
\left[ 1 - \frac{{\cal H}^{(1)}}{k_B T} 
+ \cdots \right]
\label{eq:bolt}
\end{eqnarray}
and consider only the first term on the right-hand side. 
In Ref.\onlinecite{VANV48}, the Boltzmann factor was not included 
since the temperature of interest
was quite high, so that all energy levels were equally populated. 
In our calculations, we use a mean-field approximation, so that
the sums of $A_{jk}$ and $J_{jk}$ [Eqs.~(\ref{eq:dp}) and (\ref{eq:ex})]
can be separated from the spin operators.

To compare with the measured {\it line shifts}, we calculate
perturbatively the first moment, $\langle H - H_{\mathrm res} \rangle$,
where $H_{\mathrm res}$ is the resonant field without the
spin-spin interactions,
and subtract from it the first moment at a reference temperature,
30 K. This reference temperature was chosen because at higher
temperatures additional line shifts can be expected from the temperature 
dependence of $D$.
The calculated line shifts to zero order in ${\cal H}^{(1)}/k_B T$
contain the exchange coupling constant $J$ and the effective dipole field
$\Delta \equiv \sum_{j\neq k} A_{jk}/N$ 
($N$ is the number of molecules in the sample and the summation 
runs over all molecules) as variable parameters. This zero-order
result in ${\cal H}^{(1)}/k_B T$ depends on $J$ and $\Delta$ through
the terms $({\cal E}_{n^{\prime}}-{\cal E}_n)$ in Eq.~(\ref{eq:lth_moment}). 
Since $\Delta$ depends on the sample shape due to the field-induced
net magnetization, Ewald's method\cite{EWAL21} is not sufficient 
to estimate its value in our case.

To compare with the measured {\it linewidths}, we need to
calculate the second central moment, 
$\langle (H - \langle H \rangle)^2 \rangle$, which 
is equivalent to $\langle (H - H_{\mathrm res})^2 \rangle 
- (\langle H - H_{\mathrm res} \rangle)^2$, where
$\langle (H - H_{\mathrm res})^2 \rangle$
and $(\langle H - H_{\mathrm res} \rangle)^2$ are 
calculated perturbatively to zero order in 
${\cal H}^{(1)}/k_B T$.\cite{MCMI60} 
The square root of the second central moment is proportional 
to the broadening due to the spin-spin interactions.
The quantity $\langle (H - H_{\mathrm res})^2 \rangle$ includes
the following six terms, $\sum_{ij}^{\prime} J_{ij}^2$,
$\sum_{ij}^{\prime} J_{ij} A_{ij}$,
$\sum_{ij}^{\prime} A_{ij}^2$,
$\sum_{ijk}^{\prime} J_{ij} J_{jk}$, 
$\sum_{ijk}^{\prime} J_{ij} A_{jk}$,  
and $\sum_{ijk}^{\prime} A_{ij} A_{jk}$.
The quantity $(\langle H - H_{\mathrm res} \rangle)^2$ includes
three terms, $(\sum_{ij}^{\prime} J_{ij})^2/N$, 
$(\sum_{ij}^{\prime} J_{ij})(\sum_{ij}^{\prime}A_{ij})/N$, and
$(\sum_{ij}^{\prime} A_{ij})^2/N$. Here $\sum_{ij}^{\prime}$ and 
$\sum_{ijk}^{\prime}$ run over all molecules with the constraint that
no two indices in the summations may be the same, 
and $J_{ij}$ and $J_{jk}$ are nonzero for nearest-neighbor molecules only.
The coefficients of the last three terms (the summations
over three indices $i$, $j$, and $k$) in 
$\langle (H - H_{\mathrm res})^2 \rangle$ are the same as
those of the three terms in $(\langle H - H_{\mathrm res} \rangle)^2$.
To simplify the second central moment, we use the identities
\begin{eqnarray}
\frac{1}{N} \left( \sum_{ij}\,^{\prime} J_{ij} \right)^2 
&=& \sum_{ij}\,^{\prime} J_{ij}^2 + \sum_{ijk}\,^{\prime} J_{ij} J_{jk} \:, \\
\frac{1}{N} \left( \sum_{ij}\,^{\prime} J_{ij} \right) 
\left( \sum_{ij}\,^{\prime}A_{ij} \right) 
&=& \sum_{ij}\,^{\prime} J_{ij} A_{ij} 
+ \sum_{ijk}\,^{\prime} J_{ij} A_{jk} \:,\\
\frac{1}{N} \left( \sum_{ij}\,^{\prime} A_{ij} \right)^2  
&=& \sum_{ij}\,^{\prime} A_{ij}^2 + \sum_{ijk}\,^{\prime} A_{ij} A_{jk}  \:.
\label{eq:3rd}
\end{eqnarray}
Summations over four different indices do not appear in Eq.~(\ref{eq:3rd}) 
because $\sum_{ij}^{\prime} A_{ij}=N \sum_{j\neq 1} A_{1j}$ 
by translational invariance.
We thus have only three undetermined terms, $\sum_{ij}^{\prime} J_{ij}^2$,
$\sum_{ij}^{\prime} A_{ij}^2$, and $\sum_{ij}^{\prime} J_{ij} A_{ij}$
in the calculated second central moment. The exchange 
coupling constant $J$ could be determined from the line-shift analysis. 
The remaining two terms, $\sum_{ij}^{\prime} A_{ij}^2$ and 
$\sum_{ij}^{\prime} J_{ij} A_{ij}$, can, in principle, be 
calculated from the exact geometry of the system.
However, in our study, we take the two terms as variable parameters, and 
compare their optimum values with the calculated values.
Physical justification for this is provided in Sec.~IV.A.2.

We fix the EPR frequency at $\nu=116.9$ GHz and vary the temperature
from 4~K to 35~K for the linewidth analysis (2~K to 30~K for the
line-shift analysis). We do not analyze the experimental data above 35~K
because at higher temperatures excited 
states (effective spin $S < 10$) might play a role. 
Only good-quality (high signal-to-noise ratio) experimental data were selected.
For the line-shift analysis, the exchange constant $J$
and effective dipole field $\Delta$ are varied, while for the linewidth 
analysis, $\Gamma\equiv\sum_{ij}^{\prime} A_{ij}^2/N$,
$\Lambda\equiv\sum_{ij}^{\prime} J_{ij} A_{ij}/N$, 
and the standard deviation of $D$
are varied within experimentally acceptable ranges 
in order to fit the experimental data.
Note that in Ref.\onlinecite{PARK02} the molecules of Fe$_8$ were 
assumed to be distributed on a simple cubic lattice in 
a spherical sample. For spherical samples the value of $\Delta$ 
is zero for all effective dipole distances. Thus only one parameter, 
either $\Gamma$ or the effective dipole distance, sufficed 
in the linewidth analysis.
The real samples, however, were not spherical. The Fe$_8$ sample 
examined\cite{HILL02} was a thin rhombic platelet with acute angle 
of about 60 degrees, 
edges of length 0.7~mm, and thickness 0.17~mm, and the molecules in Fe$_8$ 
are distributed on a triclinic lattice. Therefore, we here use two fitting 
parameters ($\Delta$ and $\Gamma$) to consider the dipole-dipole 
distributions in the real experimental samples described above.

\subsection{Mn$_{\mathbf 12}$}

For Mn$_{12}$ we consider an effective single-spin Hamiltonian
which satisfies tetragonal symmetry,
\begin{eqnarray}
{\cal H}_{0} &=& -D S_z^2 - C S_z^4 - g \mu_B H_z S_z 
\end{eqnarray}
with $D$$=$$0.55k_B$, $C$$=$$1.17 \times 10^{-3}k_B$, and $g$$=$$1.94$.
\cite{BARR97}
Here we consider the case in which the applied field is
along the easy axis (the $z$ axis), and we neglect the
small transverse fourth-order anisotropy term, $S_x^4+S_y^4$.

The technique is the same as for Fe$_8$, except for the 
following: (i) The $g$-strain effect provides a weak temperature
dependence to the linewidths, caused by the temperature dependence 
of the natural linewidths.
(ii) The resonant field without the spin-spin interactions is modified to 
\begin{eqnarray}
H_{\mathrm res} &\equiv& 
\frac{h \nu - D(2M_s-1) + C(4M_s^3-6M_s^2+4M_s-1)}{g \mu_B} \; .
\end{eqnarray}
(iii) To calculate the natural linewidths, we use the strength of 
the coupling between the spin system and a surrounding phonon 
heat bath from Ref.\onlinecite{LEUE00-1}.
(iv) Dipoles are distributed on a centered tetragonal lattice
with sample dimensions $1 \times 0.15 \times 0.05$~mm$^3$.
(In Ref.~\onlinecite{PARK02}, the dipoles were assumed to be distributed
on a body-centered cubic lattice, and the sample was assumed to be spherical.) 
(v) The easy anisotropy axis is along the long 
side of the needle-shaped sample. 
(vi) For Mn$_{12}$ the measured line shifts are negligible compared to
the measured linewidths, so that we do not have to consider 
the exchange interaction and the effective dipole field
($J=0$ and $\Delta=0$, so $\Lambda=0$).
Thus, the second {\it central} moment 
$\langle (H - \langle H \rangle)^2 \rangle$, which is proportional to
the measured linewidths, is identical to the second moment
$\langle (H-H_{\mathrm res})^2 \rangle$.

To compare with the experimental data,
the frequency is fixed at $\nu=189.123$ GHz, and the temperature 
is varied from 10 K to 40 K. Our analysis ends at 40 K because at higher
temperatures excited states (effective spin $S < 10$) might play a role.
\cite{HENN97,GOME98,BARB98} Only EPR spectra of good quality 
were selected, and $\Gamma$ and the 
standard deviations of $D$ and $g$ were varied 
within acceptable ranges in order to fit the experimental data. 

\section{Results and Discussion}

We show that the spin-spin interactions (dipolar and/or exchange interactions)
alone determine the trend of the temperature dependence of the line shifts
and the linewidths. From the line-shift analysis, we can estimate 
the orders of magnitude of the exchange interaction and the
effective dipole field and obtain their signs.
Using this information, we can also explain quantitatively the linewidths, 
including the $D$-strain and/or the $g$-strain effects which give rise to
a strong $M_s$ dependence but weak temperature dependence of the linewidths.
The spin-spin interactions contribute more to the linewidths 
for Fe$_8$ than for Mn$_{12}$, mainly because the
$D$-strain effect is dominant over the spin-spin interactions
for Mn$_{12}$, while it is comparable with the spin-spin interactions
for Fe$_8$. This explains the different temperature behavior of 
the linewidths for Fe$_8$ and Mn$_{12}$. 
The set of parameter values which best explains the experimental
data has some systematic theoretical uncertainties, which are difficult
to calculate exactly.

\subsection{Fe$_{\mathbf 8}$}

\subsubsection{line shifts}

\begin{figure}
\includegraphics[angle=0,width=.35\textwidth]{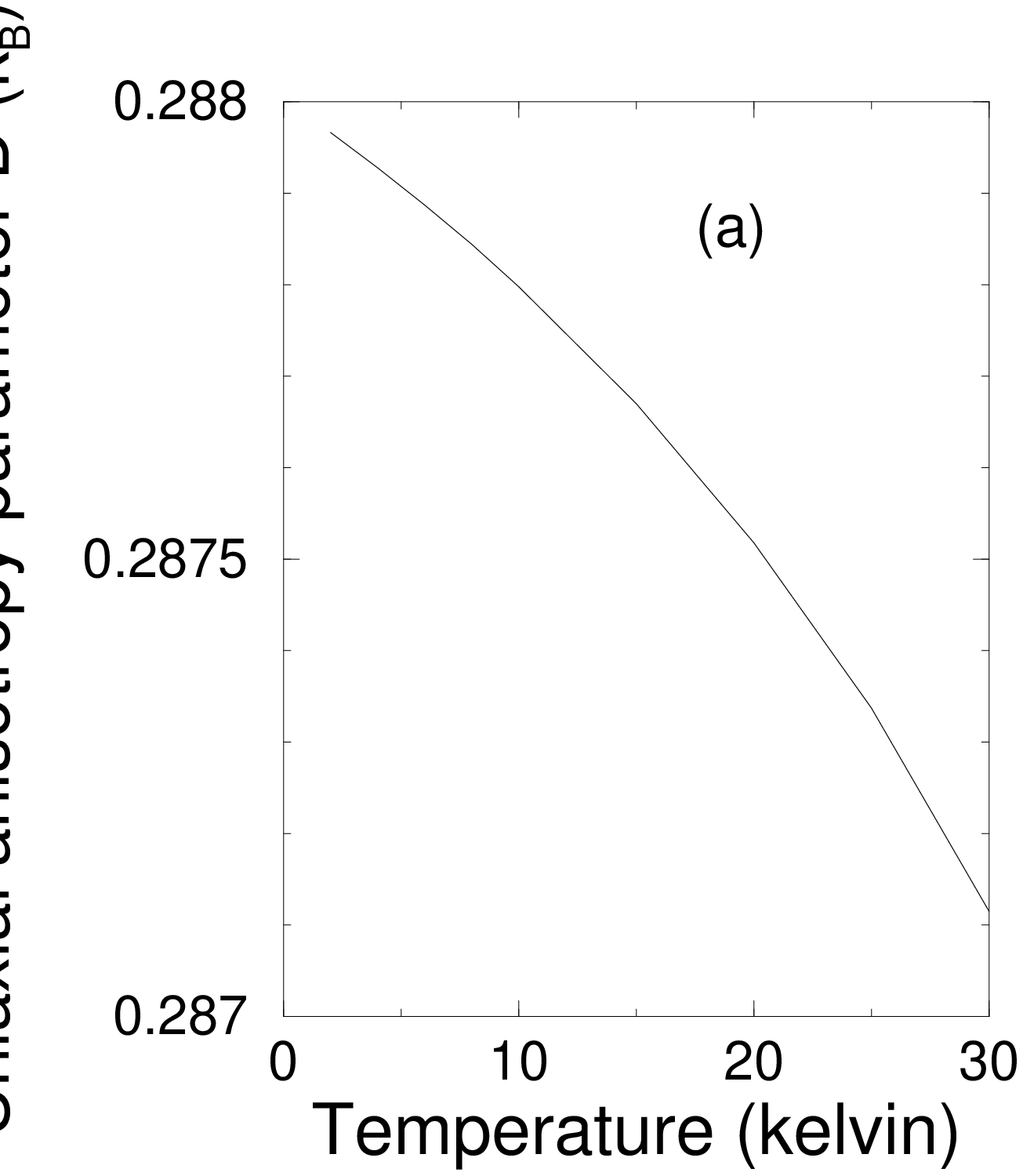}
\hspace{.2in}
\includegraphics[angle=0,width=.35\textwidth]{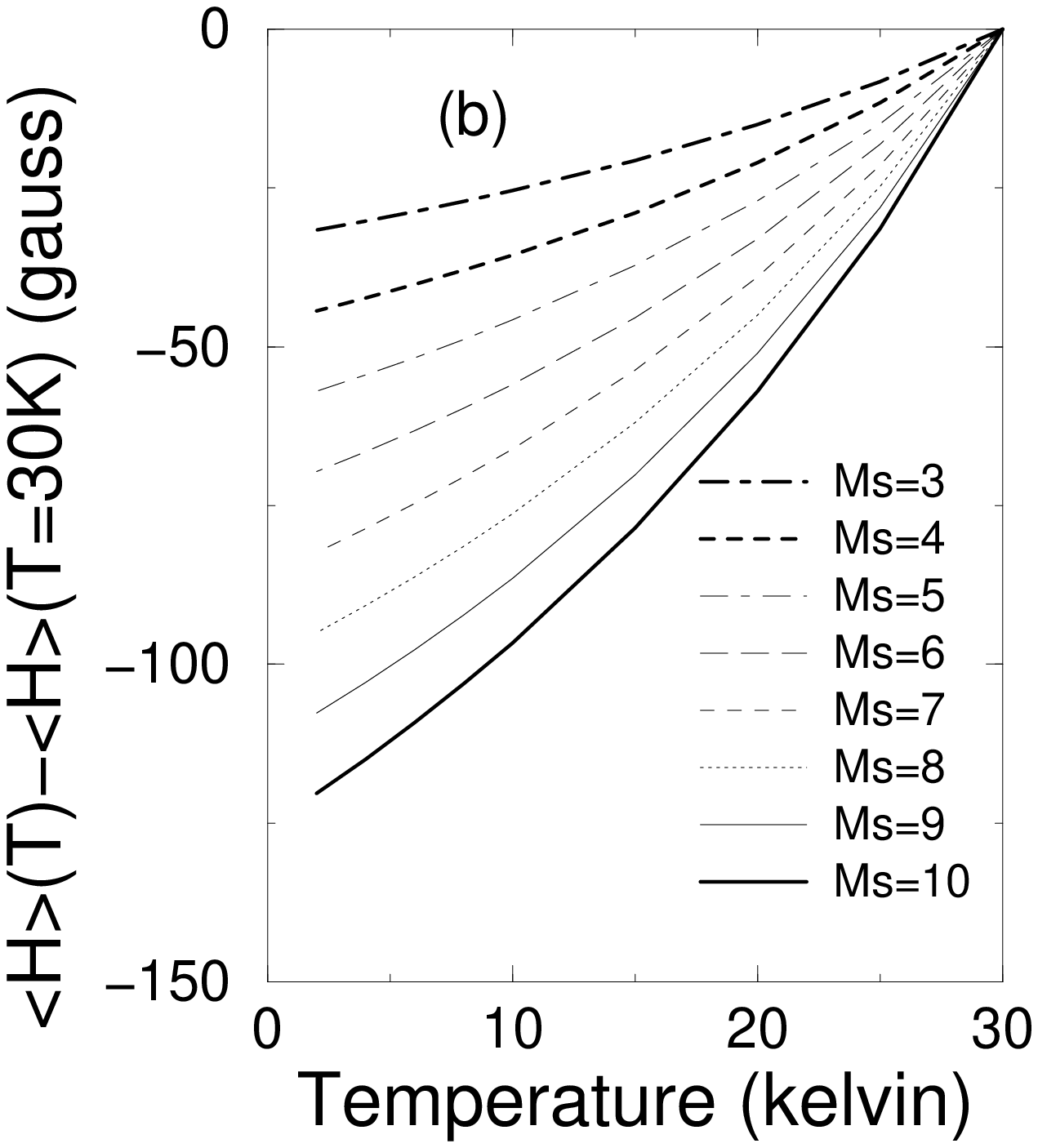}
\caption[]{\label{fig:Tdep_D}(a) Hypothetical smooth temperature 
dependence of the uniaxial anisotropy parameter 
$D$ for Fe$_8$. The functional form used here is 
$D(T)=-0.710665 + \exp\{-1/[10(50-T)+250]\}$.
(b) The resulting calculated line shifts 
(the peak position at a given temperature
minus the position at $T=30$ K) due to this temperature dependence of $D$,
shown vs temperature at $\nu=116.9$ GHz.}
\end{figure}

\begin{figure}
\includegraphics[angle=0,width=.50\textwidth]{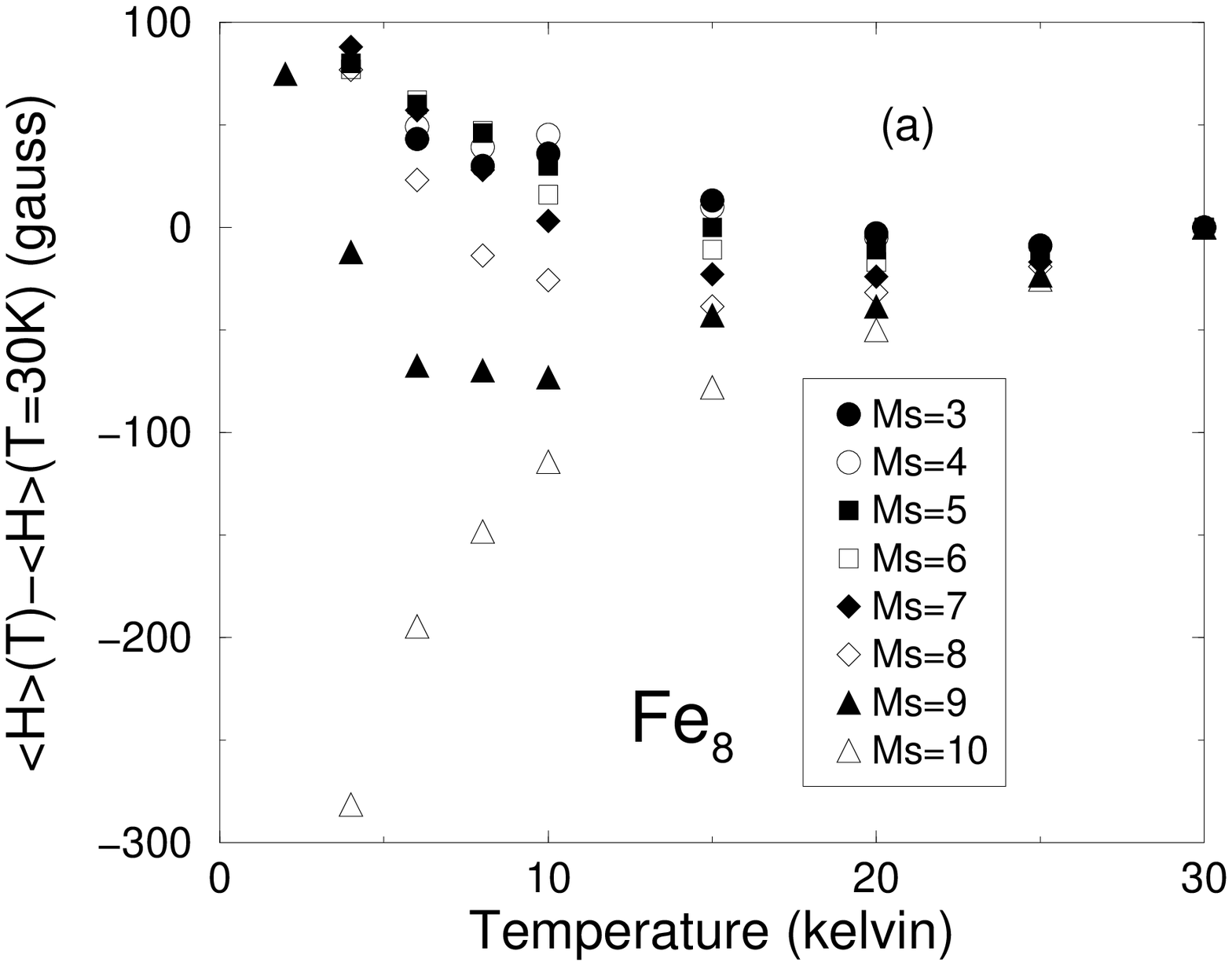}
\includegraphics[angle=0,width=.50\textwidth]{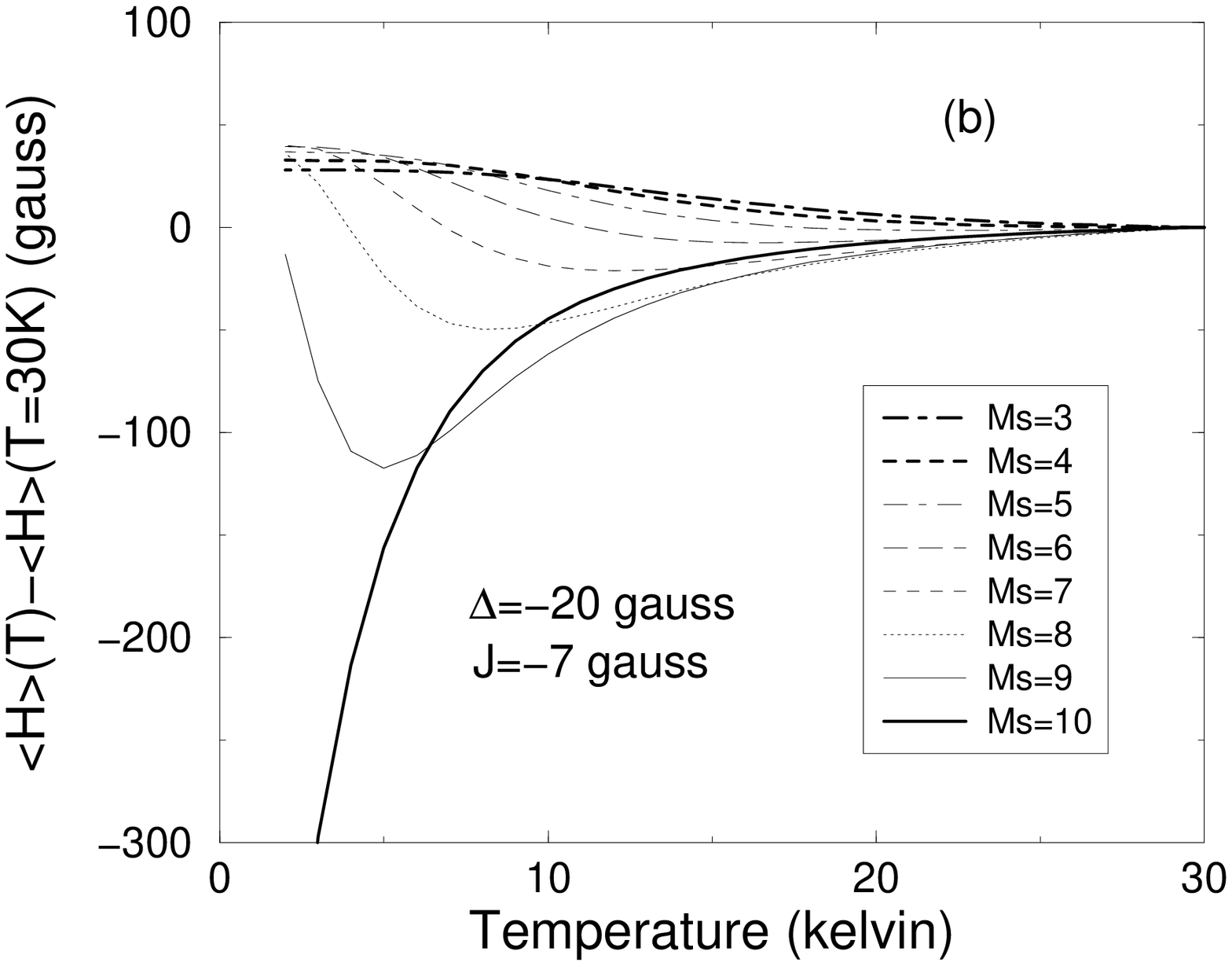}
\includegraphics[angle=0,width=.50\textwidth]{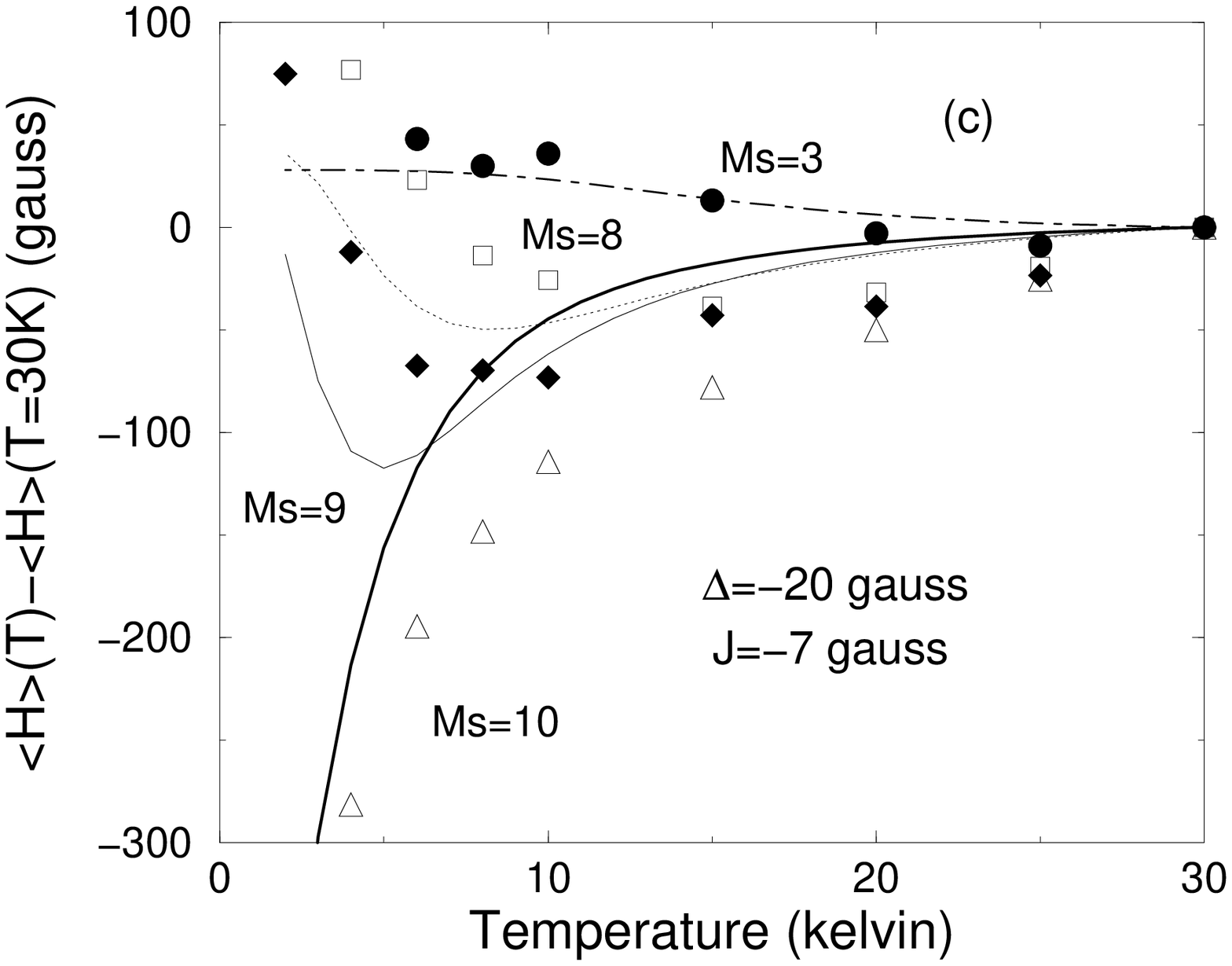}
\caption[]{\label{fig:exp_shifts}(a) Measured line shifts 
vs temperature at $\nu=116.9$ GHz for Fe$_8$. 
(b) Calculated line shifts vs temperature
at $\nu=116.9$ GHz for Fe$_8$. Here the calculation is performed
to zero order in ${\cal H}^{(1)}/k_B T$.
We use the effective dipole field 
$\Delta\equiv\sum_{ij}^{\prime} A_{ij}/N=-20$ gauss and 
the exchange coupling constant $J=-7$ gauss.
(c) Measured line shifts (symbols in (a)), superimposed
on calculated line shifts (curves in (b)),
for several transitions for comparison.}
\end{figure}

It is known that the value of the uniaxial anisotropy parameter $D$ may vary
smoothly with temperature.\cite{PENR50} To gauge the importance of
this effect, we first assume that $D$ has a temperature dependence such as 
Fig.~\ref{fig:Tdep_D}(a). Then the line shift, 
$\langle H \rangle (T)$$-$$\langle H \rangle$($T$$=$30 K),
becomes monotonically temperature dependent, 
as shown in Fig.~\ref{fig:Tdep_D}(b). Comparing Fig.~\ref{fig:Tdep_D}(b) 
with the experimental data in Fig.~\ref{fig:exp_shifts}(a), 
we see that the monotonic temperature dependence
of $D$ cannot by itself explain the complicated temperature dependence
of the measured line shifts. It could presumably contribute together with other
factors described below. However, we hereafter take
$D$ as a temperature independent parameter, for the sake of simplicity
and because the exact temperature behavior is not yet known. 
We also note that the distribution in $D$ does not change the line positions.

\begin{figure}
\includegraphics[angle=0,width=.33\textwidth]{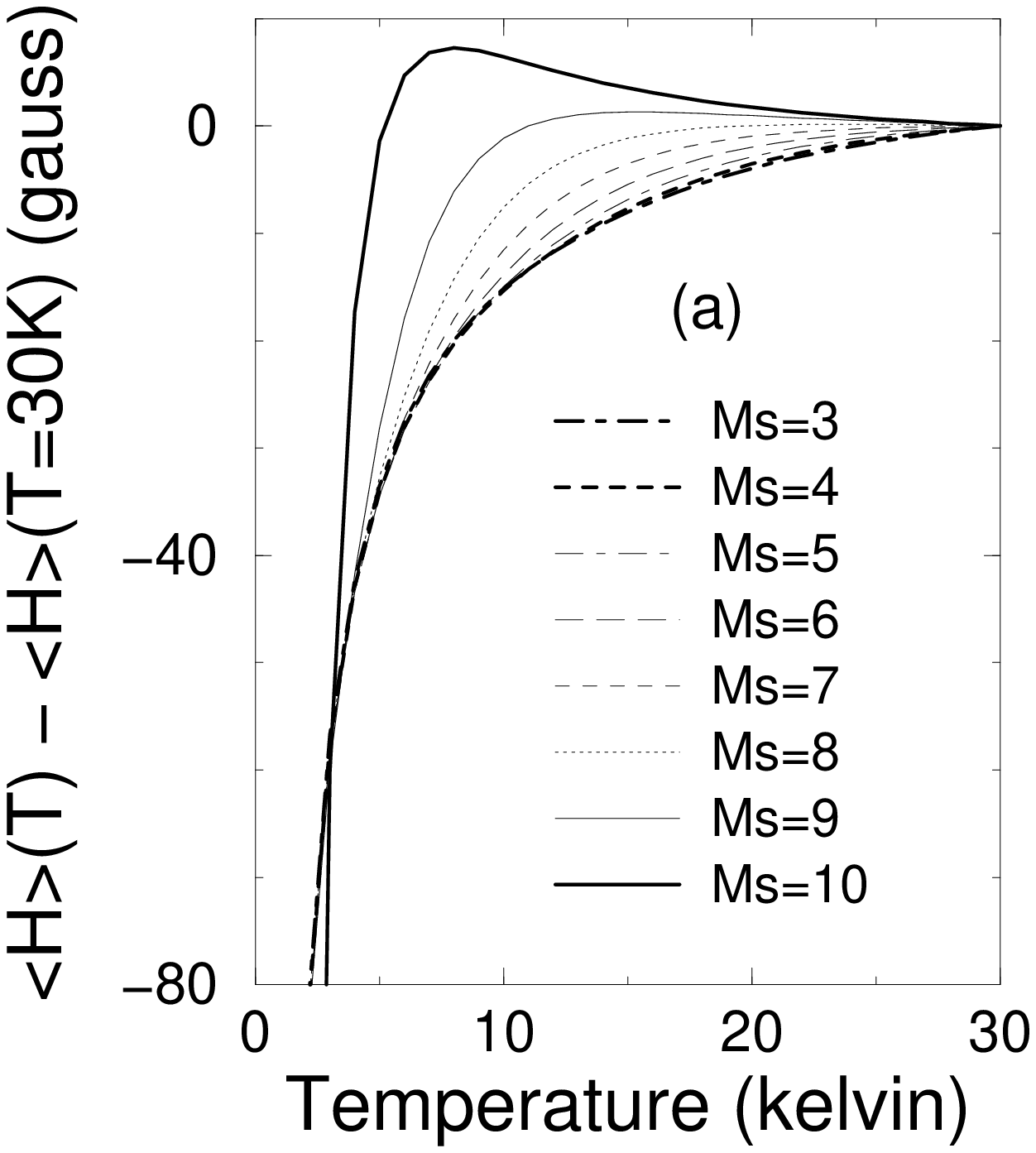}
\includegraphics[angle=0,width=.33\textwidth]{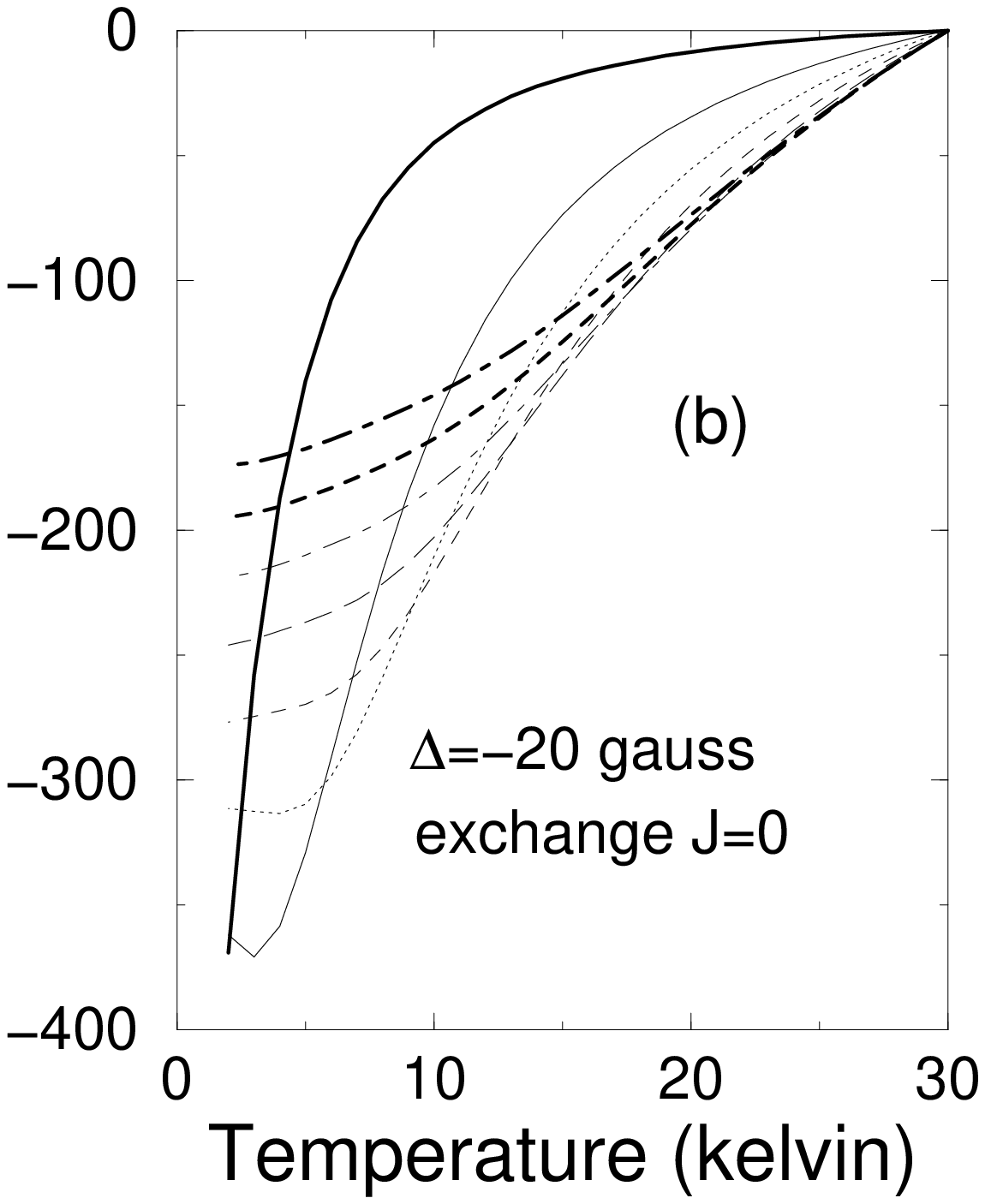}
\includegraphics[angle=0,width=.33\textwidth]{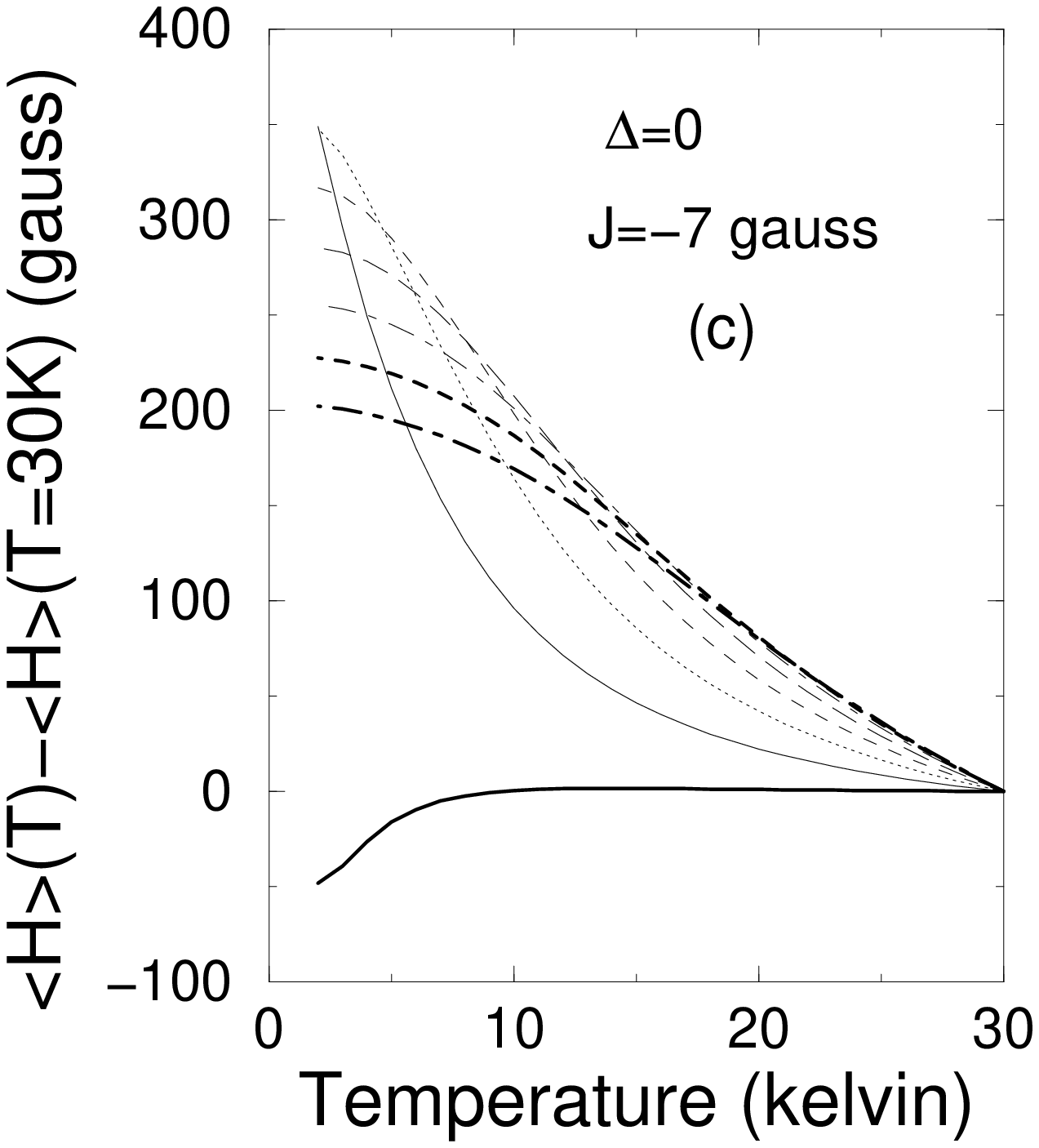}
\includegraphics[angle=0,width=.33\textwidth]{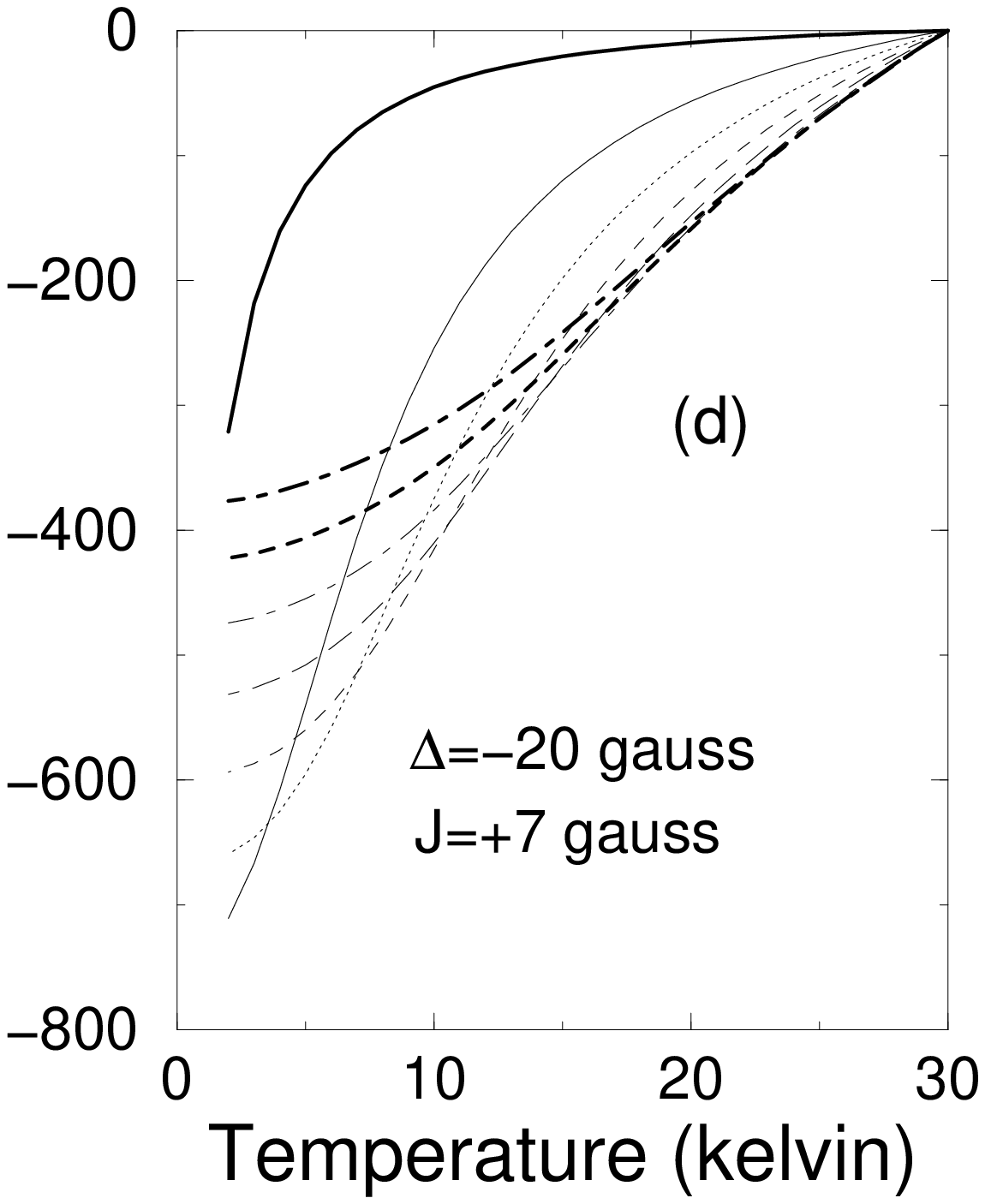}
\includegraphics[angle=0,width=.33\textwidth]{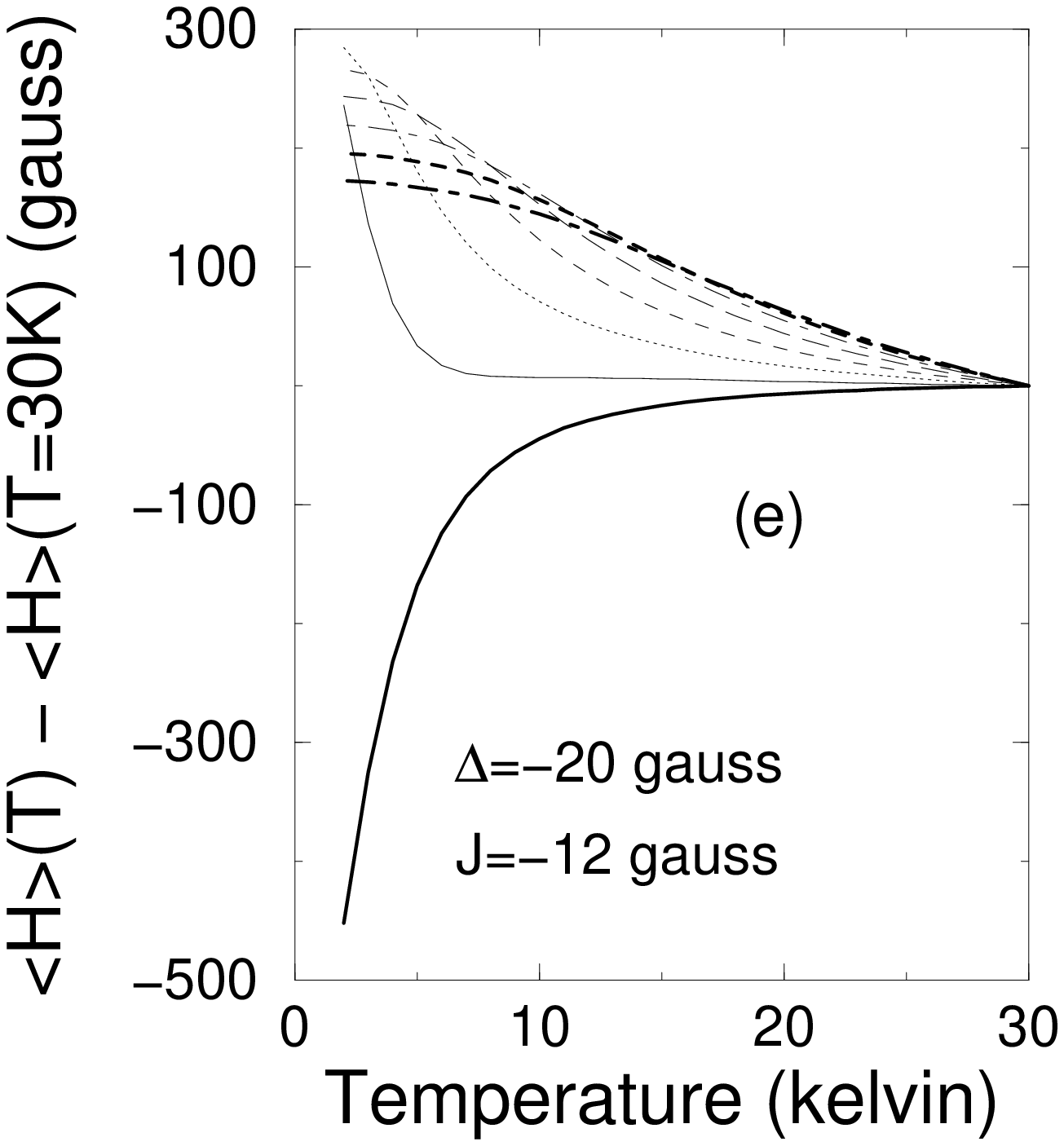}
\includegraphics[angle=0,width=.33\textwidth]{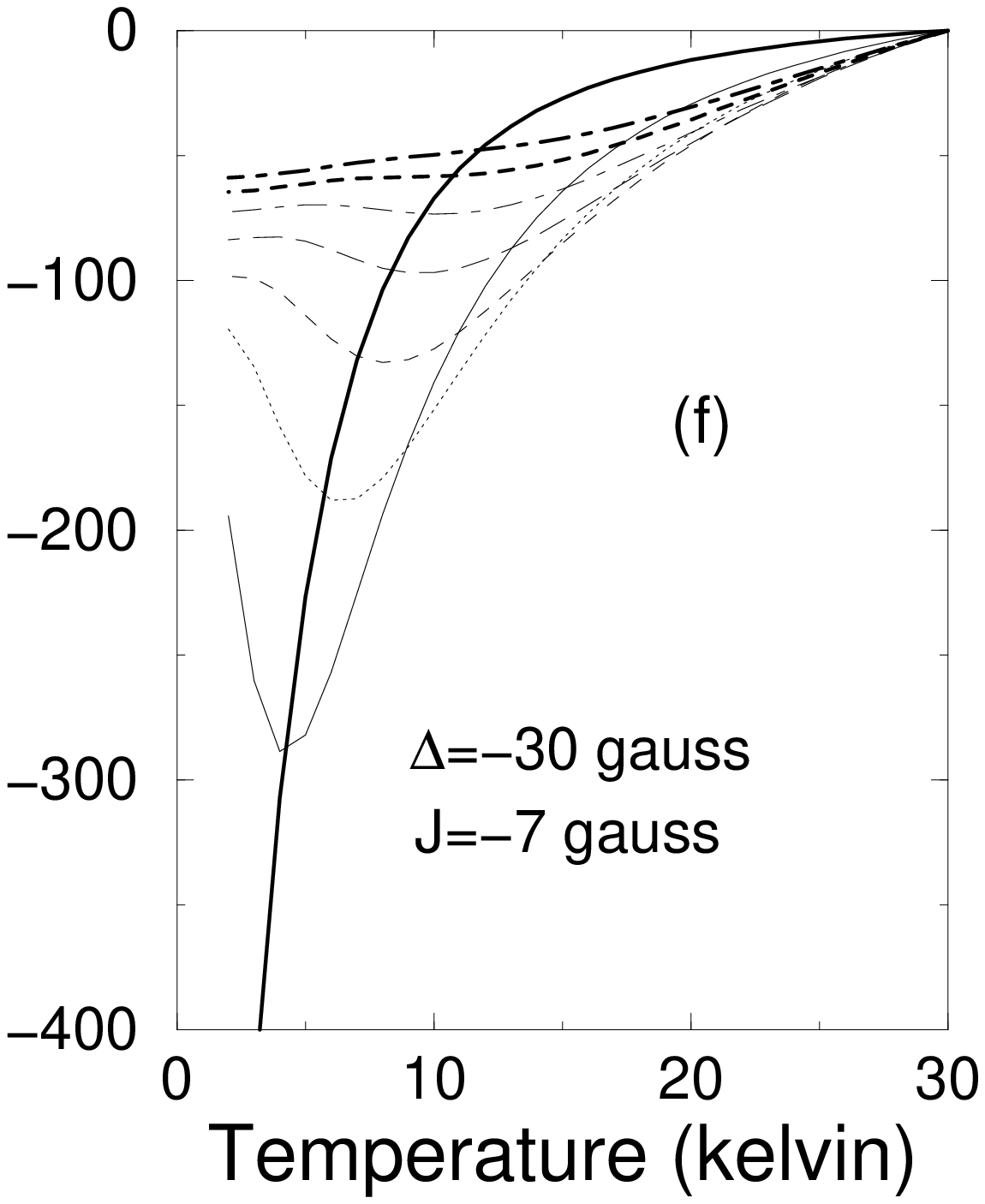}
\caption[]{\label{fig:shifts}Calculated line shifts vs temperature 
at $\nu=116.9$ GHz for Fe$_8$,
with (a) only the dipolar interaction for a spherical sample
in higher-order calculations ($J=0$, 
$\Delta\equiv\sum_{ij}^{\prime} A_{ij}/N=0$, 
and $\Gamma\equiv\sum_{ij}^{\prime} A_{ij}^2/N\neq0$) 
(b) only the effective dipole field, $\Delta=-20$ gauss, 
(c) only the exchange coupling constant, $J=-7$ gauss,
(d) $\Delta=-20$ gauss $<$ 0 and $J=+7$ gauss $>$ 0, 
(e) $\Delta=-20$ gauss and $J=-12$ gauss, and 
(f) $\Delta=-30$ gauss and $J=-7$ gauss.}
\end{figure}

Next, we consider the effect of the spin-spin interactions between
molecules on the measured line shifts.
If we first ignore the exchange interaction and consider 
the dipolar interaction only for a spherical sample with dipoles
distributed on a simple cubic lattice (which is the assumption
made in Ref.\onlinecite{PARK02}), then the effective dipole
field $\Delta$ vanishes, so that there is no line shift
to zero order in ${\cal H}^{(1)}/k_B T$. Higher-order corrections
[the second term in Eq.~(\ref{eq:bolt})] provide a 
much smaller and qualitatively different temperature dependence 
from that seen in the measurements
[compare Fig.~\ref{fig:shifts}(a) with Fig.~\ref{fig:exp_shifts}(a)].
If we include a non-zero effective dipole field $\Delta$ only
in the zero-order calculation, then
a negative effective dipole field moves the line shifts
for all the transitions down below zero [Fig.~\ref{fig:shifts}(b)].
On the other hand, if we include an exchange interaction only 
in the zero-order calculation,
then a ferromagnetic exchange interaction (negative $J$)
moves the line shifts up above zero for all transitions except 
$M_s=+10$ [Fig.~\ref{fig:shifts}(c)].
In both cases, the calculated line shifts behave very
differently from the measured shifts [Fig.~\ref{fig:exp_shifts}(a)].
Therefore, we need to include both the effective dipole field and the  
exchange interaction in order to explain the measured line shifts.
An inter-molecular exchange interaction was recently
observed for the different types of single-molecule magnets, Mn$_4$
and Mn$_4$ dimer.\cite{WERN02}
Since the effective dipole field $\Delta$ depends on the sample shape
due to the field-induced net magnetization, we do not use
Ewald's method\cite{EWAL21} to estimate $\Delta$
and thus leave it as a fitting parameter.
For the exchange interaction, we assume that the coupling constant $J_{ij}$ is 
isotropic along the $a$, $b$, and $c$ directions of the triclinic 
unit cell (although in experimental samples the exchange interactions
are highly anisotropic), so that the coordination number is 6.  
The optimum values of $\Delta$ and $J$ are 
$\Delta_{\mathrm opt} \approx -20$ gauss 
and $J_{\mathrm opt} \approx -7$ gauss ($\sim$ 1 mK) respectively. 
With these optimum values, the calculated line shifts 
[Fig.~\ref{fig:exp_shifts}(b)] reproduce well the trends of
the temperature dependence of the experimental data 
[Fig.~\ref{fig:exp_shifts}(a)]. Figure~\ref{fig:exp_shifts}(c) 
shows a direct comparison between theory and experiment
for a few transitions. 

The negative sign of the effective dipole field ($\Delta <0$) 
indicates that dipoles are antiferromagnetically coupled.
This result seems to be in conflict with the prediction that
the dipolar Ising spin system with the same structure as Fe$_8$ 
is ferromagnetically ordered.\cite{FERN00,MART01} 
However, as pointed out in Ref.~\onlinecite{MART01}, 
the energy difference between the ferromagnetic and antiferromagnetic
states is so small that any neglected effects may shift the ground
state to an antiferromagnetic state. 
The negative sign of the exchange coupling constant $(J < 0)$ 
corresponds to ferromagnetic coupling between the 
effective spins of the molecules.
Thus, if there exists any ordering for Fe$_8$, then the ordering 
temperature should be estimated by considering both the exchange and dipolar 
interactions. The two interactions compete with each other, thereby reducing
the possible ordering temperature to a lower value 
than the ordering temperature 
with only one of the two interactions considered. 

Finally, we show the calculated line shifts with several other
parameter values that are different from the optimum ones.
If the effective dipole field and the exchange interaction 
both change signs, then the calculated line shift also
changes its sign. If the sign of $\Delta$ is opposite to
the sign of $J$, then the magnitude of
the calculated line shift for $M_s=+10$ is much smaller than
those for the other transitions, which does not agree with 
the experimental data [compare Fig.~\ref{fig:shifts}(d) with
Fig.~\ref{fig:exp_shifts}(a)].
Figure~\ref{fig:shifts}(e) shows the calculated line shifts with
$J < J_{\mathrm opt}$ and $\Delta=\Delta_{\mathrm opt}$.
Figure~\ref{fig:shifts}(f) shows the calculated line shifts 
with $J=J_{\mathrm opt}$ and $\Delta < \Delta_{\mathrm opt}$.
All three figures [Figs.~\ref{fig:shifts}(d)-(f)] are significantly
different from Fig.~\ref{fig:exp_shifts}(b) with the optimized values.

\subsubsection{linewidths}

\begin{figure}
\includegraphics[angle=0,width=.50\textwidth]{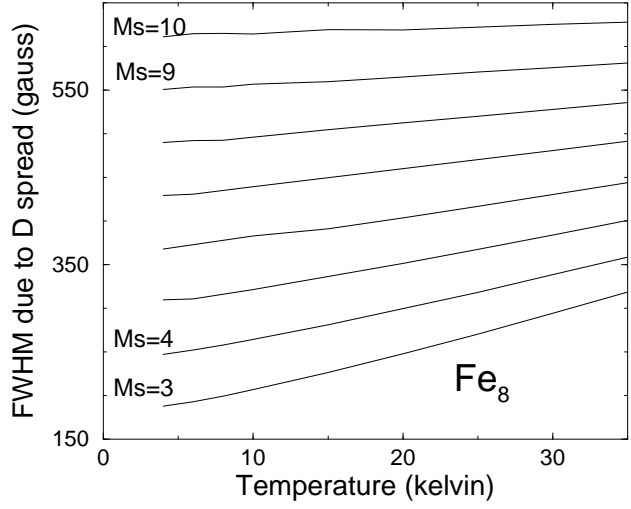}
\caption[]{\label{fig:D}Calculated full width at half maximum (FWHM)
caused by the Gaussian distribution in the uniaxial
anisotropy parameter $D$ only, shown vs
temperature from 4 K to 35 K at $\nu=116.9$ GHz for Fe$_8$.
$M_s=10$ indicates the transition from the energy level $M_s=10$ to 
$M_s=9$, etc. The standard deviation of $D$ is approximately $0.0064D$.}
\end{figure}

For Fe$_8$ the distribution in $D$ and the spin-spin interactions
contribute approximately equally to the inhomogeneous line broadening.
Figure~\ref{fig:D} shows the calculated line broadening due to 
the $D$-strain effect only as a function of temperature at $\nu=116.9$ GHz. 
Here the standard deviation of the Gaussian distribution in $D$,
$\sigma_D$, is approximately $0.0064D$. 
The line broadening caused by the $D$-strain only becomes temperature
dependent, because the natural linewidths depend on temperature.
The distribution in $D$ makes each molecule subject to a slightly
different resonant field. A measured line shape is a sum of
many Lorentzian line shapes with a natural linewidth and different
resonant fields. We can calculate the variance of the
resonant field, $\sigma_D (2M_s-1)/g \mu_B$, due to the distribution 
in $D$ from the expression for $H_{\mathrm res}$. 
If the natural linewidths are comparable with the variance of the 
resonant field, then the effect of temperature is significant.
If the natural linewidths are much smaller than the variance, 
then the effect of temperature is negligible.
The natural linewidth at 10 K (35 K) 
varies from 7 G (29 G) to 79 G (235 G) as $M_s$ changes from +10 to +3,
using the parameter values in Ref.\onlinecite{LEUE00-2}.
The variance of the resonant field,
$\sigma_D (2M_s-1)/g \mu_B$, varies from 260 G to 70 G as 
$M_s$ is varied from +10 to +3 with $\sigma_D=0.0064D$.
Thus, we find that for small $M_s$ the natural linewidths
are comparable with the variance, while for large $M_s$
the natural linewidths are much smaller than the variance.
Therefore, for small $M_s$ the calculated linewidths 
show a substantial temperature 
dependence, while for large $M_s$ there is only a very weak 
temperature dependence (see Fig.~\ref{fig:D}).

\begin{figure}
\includegraphics[angle=0,width=.50\textwidth]{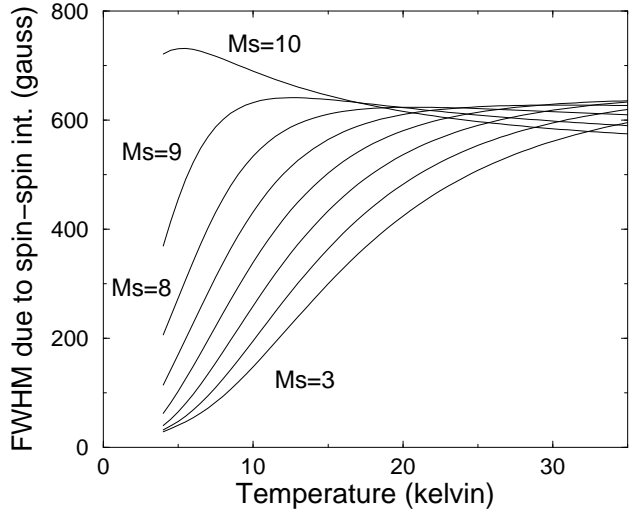}
\caption[]{\label{fig:dp}Calculated FWHM caused by the spin-spin 
interactions only, shown vs temperature at $\nu=116.9$ GHz for Fe$_8$.
Here the exchange constant $J$ is $-7$ gauss, 
$\Gamma\equiv\sum_{ij}^{\prime} A_{ij}^2/N=86$ gauss$^2$ and
$\Lambda\equiv\sum_{ij}^{\prime} J_{ij} A_{ij}/N = -156$ gauss$^2$.}
\end{figure}

\begin{figure}
\includegraphics[angle=0,width=.50\textwidth]
{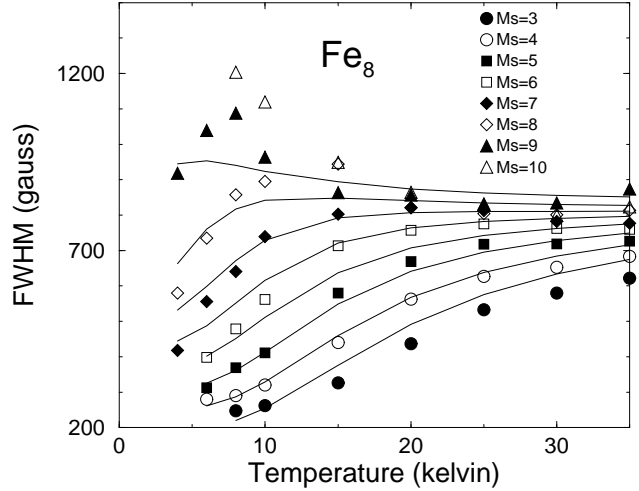}
\caption[]{\label{fig:D+dp}Calculated (curves) and measured FWHM (symbols) vs 
temperature at $\nu=116.9$ GHz for Fe$_8$. Here we use
the standard deviation of $D$, $\sigma_D \approx 0.0064D$, 
the exchange coupling constant, $J= -7$ gauss, $\Gamma=86$ gauss$^2$,
and $\Lambda=-156$ gauss$^2$. The solid curves, from bottom to
top, correspond to $M_s=3,4, ...,9,10$.
See the text for possible sources of the discrepancy between theory
and experiment for low temperatures and large $M_s$.}
\end{figure}

In Fig.~\ref{fig:dp} the calculated line broadening caused solely by 
the spin-spin interactions at fixed $D$ is shown vs temperature 
at $\nu=116.9$ GHz. Here we use the exchange constant, $J=-7$ gauss,
which was estimated from the measured line shifts (Sec.~IV A.1),
$\Gamma\equiv\sum_{ij}^{\prime}A_{ij}^2/N=86$ gauss$^2$, and
$\Lambda\equiv\sum_{ij}^{\prime} J_{ij} A_{ij}/N=-156$ gauss$^2$.
For the ground state $M_s=+10$, the linewidths decrease with 
increasing temperature in the whole examined temperature range. 
For $M_s=+9,+8$, and +7, the widths first increase sharply with 
temperature at low temperatures, and then decrease slowly 
with temperature at high temperatures. For $M_s=+6,+5,+4$, and +3, 
the widths increase with increasing temperature in the whole 
examined temperature range.
As the temperature increases, the $M_s$ dependence of the line
broadening due to the spin-spin interactions decreases.
This trend was also seen in the experimental linewidths 
(shown as symbols in Fig.~\ref{fig:D+dp}), confirming that 
the spin-spin interactions are essential to understanding the 
temperature dependence of the linewidths. 

The trends of the temperature dependence 
can be qualitatively understood through the
relative magnitude difference of the thermal energy and 
the Zeeman energy splitting between the states $M_s=+10$ and $M_s=-10$.
If the Zeeman energy splitting is much larger than the thermal
energy (this occurs at low temperatures), 
then the system is polarized. Thus, higher temperature
provides larger populations in higher energy levels within the same 
potential well where the ground state is located.
This leads to an increase in the randomness of the spin orientation
so that linewidths become larger with increasing temperature. 
If the Zeeman energy splitting is much smaller than the thermal energy
(this occurs at high temperatures), then some energy levels in 
{\it both} potential wells are populated. In this case, 
thermal fluctuations increase rapidly with increasing temperature,
so that the duration time of the local magnetic field due to
neighboring molecules becomes shorter than the spin-spin relaxation time
$T_2$ (the inverse of the natural linewidths).
Eventually, at very high temperatures 
the local field is averaged out. Therefore, the linewidths 
decrease with increasing temperature,
which usually occurs in paramagnetic materials with very small
or zero single-ion anisotropy.\cite{LARS89} (This effect is
called motional narrowing.\cite{AABB})
Therefore, the ``crossover'' temperature 
where the maximum of the linewidth occurs must be
proportional to the Zeeman energy splitting between $M_s=+10$
and $M_s=-10$, which is $2 g \mu_B H_{\mathrm res} \cdot 10$.
For example, for the transition $M_s=+10 \rightarrow +9$ at $\nu=116.9$ GHz, 
the resonant field is less than 0.1 T, so the Zeeman splitting
is about 2$k_B$. Consequently, its crossover temperature is below the examined
temperature range. The crossover temperature increases with
decreasing $M_s$ because the resonances are observed at increasing fields
for decreasing $M_s$. For the transitions 
$M_s=+9 \rightarrow +8, +8 \rightarrow +7$,
and $+7 \rightarrow +6$, the crossover temperatures are
within the examined temperature range
(see the inset in Fig.~6 of Ref.\onlinecite{HILL02}).
For $M_s=+6,+5,+4$, and +3, the crossover temperatures are above the
studied temperature range. 

Figure~\ref{fig:D+dp} shows the experimental data (symbols), and 
our calculated linewidths (curves), including both the $D$-strain 
effect and the spin-spin interactions with $\sigma_D \approx 0.0064D$,
$J=-7$ gauss, $\Gamma=86$ gauss$^2$, and $\Lambda=-156$ gauss$^2$.
(The spread in $D$ here is different from that reported in 
Ref.\onlinecite{PARK02}, because $\sigma_D$ is sample dependent
and the samples examined were different. 
The value of $\Gamma$ corresponding to an effective dipole
distance of 12 \AA~(Fe$_8$) in Ref.\onlinecite{PARK02} was about 
203 gauss$^2$.)
As shown in Fig.~\ref{fig:D+dp}, our calculated linewidths
agree well with the experimental data, except in the low-temperature 
range for large $M_s$ transitions ($M_s=+10,+9,+8$).
The experimental linewidths for $M_s=+10$ show $1/T$ dependence
in the whole examined temperature range.\cite{PPHMF}
For $M_s=+10,+9$, and +8, the calculated linewidths are appreciably
smaller than the experimental linewidths below 10 K.
As a possible explanation for this discrepancy, 
we speculate that at low temperatures and large $M_s$, 
(i) our assumption, 
${\cal H}^{(1)}/k_B T \ll 1$, may break 
down, and/or (ii) there might be other linewidth contributions that
we have neglected, which should be included along with the dipolar and
exchange interactions.
In principle, when 
${\cal H}^{(1)}/k_B T$ is not much
smaller than unity, a first-order calculation 
in ${\cal H}^{(1)}/k_B T$
produces corrections of ${\cal O}(1/k_B T)$. But its implementation 
is quite complicated, and a first-order calculation may anyway not 
be sufficient to explain fully the measured linewidths.

Introducing the concept of the crossover temperature to explain
the temperature dependence of the widths 
seems to be successful for $\nu=116.9$ GHz.
However, recent EPR experiments (Fig. 8 in Ref.\onlinecite{HILL02})
showed that even when the frequency increased to $\nu=145.9$ GHz 
such that the resonant field for the ground state transition ($M_s=10$) 
is approximately 1 tesla, the linewidths for this transition still 
increased with {\it decreasing} temperatures down to 2 K. 
This cannot be explained using the reasoning given above,
because the crossover temperature for the transition is 
approximately 15 K, so that the linewidths should decrease with decreasing 
temperature below about 15 K. At present, we do not fully
understand the broadening of this ground-state transition.
(At $\nu=145.9$ GHz, for other transitions than the ground-state transition,
the temperature dependence of the linewidths can be understood using
the concept of the crossover temperature.)

Consideration of exchange interaction in the linewidths slightly reduces
$\sigma_D$ (from 0.0076D to 0.0064D)
and the dipolar interaction (from $\Gamma=103$ gauss$^2$ to $\Gamma=86$
gauss$^2$). However, the quality of the linewidth fit including
exchange interaction is comparable to that without exchange interaction
since the exchange coupling constant is very small compared with the 
linewidths.\cite{AABB} The two fitting parameters, $\Gamma=86$ gauss$^2$, 
and $\Lambda=-156$ gauss$^2$, can be calculated using the exact geometry of 
the system. The calculated values are $\Gamma_{\mathrm cal}=500$ gauss$^2$ 
and $\Lambda_{\mathrm cal}=-137$ gauss$^2$, when the easy axis
is 9$^{\circ}$ off from the $a$ axis toward the positive $b$ axis,
and 7$^{\circ}$ off from the $ab$ plane.\cite{WERN02-2}
The optimum value of $\Lambda$ 
is quite close to $\Lambda_{\mathrm cal}$, in contrast with $\Gamma$.
Possible reasons that the optimum value of $\Gamma$ is much smaller than 
$\Gamma_{\mathrm cal}$ are as follows: (i) In our calculation,
we considered each molecule to be a point dipole. If we consider the atomic
positions of the eight Fe ions in each molecule and calculate the dipolar
interaction between Fe ions in different molecules, then 
the sum of the squared dipolar interaction, $\Gamma$,
can be significantly reduced. (ii) Recent NMR experiments for 
the single-molecule magnet Mn$_{12}$ showed some spin-density 
leakage onto the ligands.\cite{ACHE01-1,ACHE01-2,ACHE02} 
This indicates indirectly that for Fe$_8$ the spin density 
in a single molecule may not be confined only on the core, 
which would thus reduce the magnetic moments of the eight Fe ions.
The above two reasons are our speculations to explain the 
discrepancy, but it is still unclear
why considering the atomic structure within each molecule
does not substantially change the value of $\Lambda$.

\subsection{Mn$_{\mathbf 12}$}

The experimental data for Mn$_{12}$ are limited
to resonance frequencies below 190 GHz, so that the large $M_s$ 
transitions where
the line shifts are significant cannot be observed for these
low-frequency measurements. Additionally, the linewidths
for Mn$_{12}$ are an order of magnitude larger than those for
Fe$_8$. Therefore, relatively small line shifts are probably
masked for Mn$_{12}$. Thus, hereafter, the small line shifts are ignored 
in our analysis, so the exchange interaction
and the effective dipole field $\Delta$ need not be 
considered in the linewidth analysis for Mn$_{12}$.
The sources of the line broadening are then $D$-strain, $g$-strain,
and the dipolar interaction.

\begin{figure}
\includegraphics[angle=0,width=.50\textwidth]
{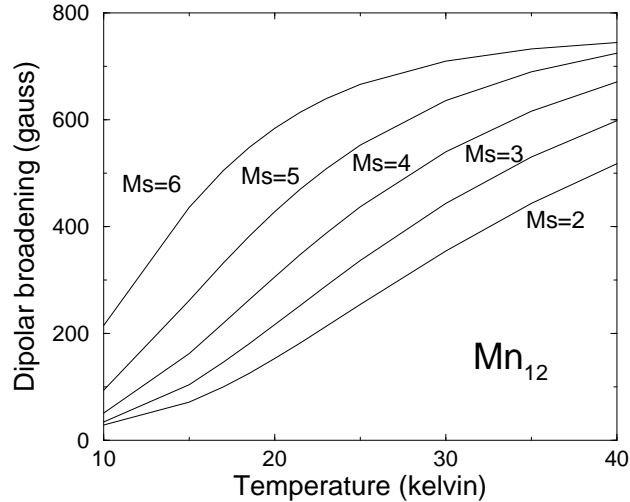}
\caption[]{\label{fig:dp-Mn}Calculated FWHM for Mn$_{12}$ caused by 
the dipolar interactions only,
shown vs temperature, at $\nu=189.123$ GHz, with the sum of
the squared dipolar interactions $\Gamma=203$ gauss$^2$.
The examined temperature range for Mn$_{12}$ is from 10 K to 40 K.}
\end{figure}

\begin{figure}
\includegraphics[angle=0,width=.50\textwidth]{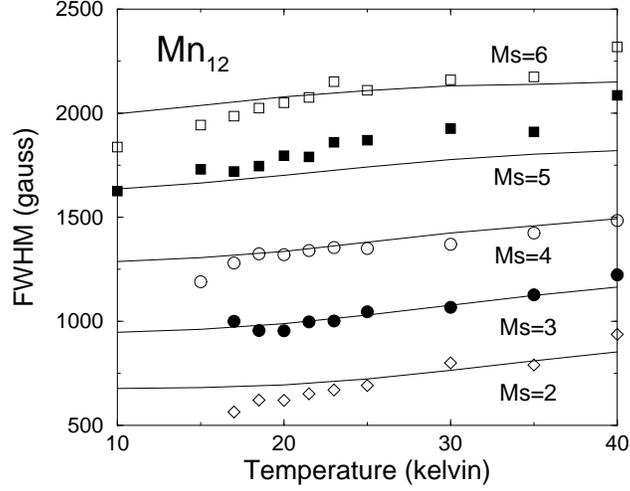}
\caption[]{\label{fig:D+g+dp}Calculated (curves) and measured (symbols) FWHM
vs temperature at $\nu=189.123$ GHz for Mn$_{12}$. Here 
the $D$-strain ($\sigma_D \approx 0.018D$), $g$-strain
($\sigma_g \approx 0.002g$), and the dipolar interactions 
($\Gamma \approx 203$ gauss$^2$) are included in the calculated linewidths.}
\end{figure}

Line broadening caused by the $D$-strain and $g$-strain effects 
for Mn$_{12}$ is found to have a weak temperature dependence (not shown),
which is similar to the line broadening due to the $D$-strain effect
for Fe$_8$. The contribution of the dipolar interaction to the linewidths
is shown vs temperature in Fig.~\ref{fig:dp-Mn}. 
Here $\Gamma\equiv\sum_{ij}^{\prime} A_{ij}^2/N=203$ gauss$^2$.
The dipolar broadening increases with increasing temperature for 
$M_s=+6,+5,+4,+3$, and $+2$. We do not see the regime where
the dipolar broadening decreases with increasing temperature,
because the crossover temperature for $M_s=+6$, about 32 K 
(the resonant field is about 1.6 T), 
is close to the highest temperature analyzed (40 K).
Unlike Fe$_8$, the $M_s$ dependence of the dipolar broadening 
does not decrease with increasing temperature 
(the curves are almost parallel). This is
also observed in the experimental data (shown as symbols in
Fig.~\ref{fig:D+g+dp}). 

We combine the three effects ($D$-strain, $g$-strain, and 
dipolar interactions) to find that the calculated linewidths
agree well with the measured linewidths with $\sigma_D \approx 0.018D$,
$\sigma_g \approx 0.002g$, and $\Gamma=203$ gauss$^2$,
as shown in Fig.~\ref{fig:D+g+dp}. (The value of $\Gamma$ 
corresponding to an effective dipole distance of 14 \AA~ for Mn$_{12}$
in Ref.\onlinecite{PARK02} turned out to be the same as that obtained
for Fe$_8$ in Ref.~\onlinecite{PARK02}.)
Here the standard deviation of $g$ is quite small, so that we cannot rule out
the possibility of $\sigma_g=0$. Note that $\sigma_D$ and $\sigma_g$ 
vary from sample to sample. The optimum parameter values 
found here are different from those estimated in Ref.\onlinecite{PARK02}, 
because the examined Mn$_{12}$ sample was different.
The calculated value for $\Gamma$, with the exact geometry of 
Mn$_{12}$ from Ref.\onlinecite{LIS80} (with each molecule considered
as a point dipole), is $\Gamma_{\mathrm cal}=397$ gauss$^2$;
this is, again, quite a bit higher than 
the optimum value for $\Gamma$, probably for the same reasons as for Fe$_8$
although the origin of this discrepancy remains unclear.
Overall, the temperature dependence 
of the linewidths for Mn$_{12}$ is weaker than
for Fe$_8$, because the distribution in $D$ for Mn$_{12}$ is 
roughly three times as wide as for Fe$_8$, and the dipolar
broadening for Fe$_8$ is comparable to that for Mn$_{12}$.
Thus, the distribution in $D$ conceals the significant 
temperature dependence of the dipolar broadening for Mn$_{12}$.

\begin{figure}
\includegraphics[angle=0,width=.50\textwidth]{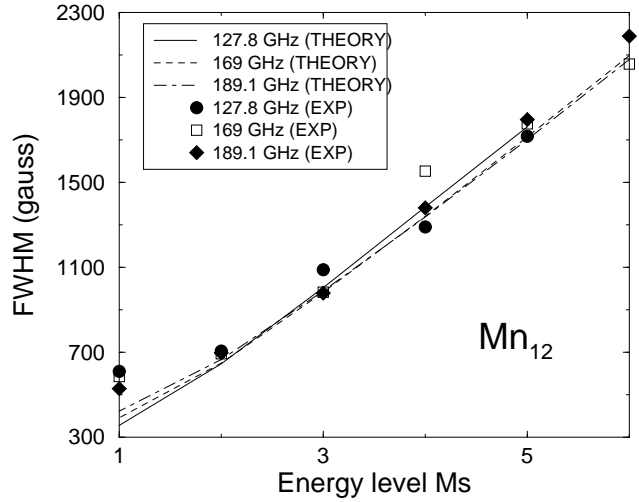}
\caption[]{\label{fig:Mn12_Ms}Calculated (curves) and measured (symbols) FWHM
vs energy level $M_s$ at $T=20$ K for $\nu=127.8$, $169$, and $189.1$ GHz 
for Mn$_{12}$. Here the values of $\sigma_D$, $\sigma_g$, and $\Gamma$
are the same as those in Fig.~\ref{fig:D+g+dp}. Because of
a relatively small contribution of the dipolar interaction to
the linewidths, the linewidths do not change much 
with the resonance frequency.}
\end{figure}

As a consistency check, we also used the same values of the three parameters 
($\sigma_D \approx 0.018D$, $\sigma_g \approx 0.002g$, 
and $\Gamma=203$ gauss$^2$)
to analyze the measured linewidths\cite{HILL02} as 
functions of the energy level $M_s$ for several frequencies
($\nu=127.8, 148.5, 169, 181.8$, and 189.1 GHz) at a fixed temperature
(T=20 K). Our calculated linewidths 
are in good agreement with the experimental data as shown
in Fig.~\ref{fig:Mn12_Ms}. Due to a dominant contribution of
the distribution in $D$ to the linewidths, the linewidths
do not depend much on the resonance frequency.

\section{Conclusion}

We have investigated how the EPR line shifts and linewidths 
vary with temperature for different energy levels $M_s$ 
with the applied field along the easy axis 
for the single-molecule magnets Fe$_8$ and Mn$_{12}$.
Our calculations consider the spin-spin interactions between molecules, 
as well as distributions in $D$ and $g$. We have found that the distributions 
in $D$ and $g$ provide a weak temperature dependence to 
the linewidths, and that the spin-spin interactions (exchange
and dipolar interactions) dominate the temperature 
dependence of the line shifts and the linewidths. 
For Fe$_8$, the line-shift analysis (Figs.~\ref{fig:exp_shifts} and
\ref{fig:shifts}) provides possible evidence
of an exchange interaction between molecules, and it determines
the sign and order of magnitude of the exchange interaction.
The competition of the suggested exchange interaction with the dipolar 
interaction would tend to lower a possible magnetic ordering temperature.
A small exchange interaction does not affect the linewidth analysis
significantly because the exchange coupling constant is much smaller than
the typical linewidths. Table~\ref{tab:opt} summarizes 
the optimized values of the parameters used in our analysis.
Those parameters are, in principle, independent of the resonance
frequencies, but some of them are expected to be somewhat 
batch and shape dependent, as indicated in Table~\ref{tab:opt}.
Because of the much broader distribution in $D$ for Mn$_{12}$, 
the linewidths for Fe$_8$ show a stronger temperature 
dependence than those for Mn$_{12}$. This conclusion also 
corroborates our assumption that $D$ is distributed 
for both materials,\cite{PARK02} although the microscopic origin of this
spread is not yet well understood.\cite{CHUD01,CORN01}


\section*{Acknowledgments}

This work was funded by NSF Grant Nos.~DMR-9871455, DMR-0120310,
DMR-0103290, and DMR-0196430, Research Corporation (S.H.), 
and by Florida State University through the School of 
Computational Science and Information Technology and
the Center for Materials Research and Technology.

\clearpage

\begin{table}[ht]
\caption[]{\label{tab:opt}Optimum values of the parameters used in the
line-shift and linewidth analysis. Here $\sigma_D$ is the standard
deviation of $D$, $\sigma_g$ is the standard deviation of $g$, $J$ is the
exchange coupling constant between nearest-neighbor molecules (negative sign
means ferromagnetic interaction), $\Delta\equiv\sum_{ij}^{\prime}A_{ij}/N$
which vanishes for spherical samples, 
$\Gamma\equiv\sum_{ij}^{\prime} A_{ij}^2/N$, and
$\Lambda\equiv\sum_{ij}^{\prime} J_{ij} A_{ij}/N$. 
With experimentally determined values of $D$ and $g$,\cite{BARR97,MACC01}
for Fe$_8$ we optimize $J$ and $\Delta$ for the line shifts and
$\sigma_D$, $\Gamma$, and $\Lambda$ for the linewidths,
while for Mn$_{12}$ we optimize $\sigma_D$, $\sigma_g$, and $\Gamma$
for the linewidths. The parameters are essentially independent of the
measurement frequencies, but some of them are expected to be 
somewhat batch and shape dependent, as marked by X in the table below.}
\begin{ruledtabular}
\begin{tabular}{c l  l  c  c  c  c  c}
 & Fe$_8$ & Mn$_{12}$ & & Batch & Shape & Size & 
Crystal structure 
\\ \hline
$D$ & 0.288 $k_B$ & 0.55 $k_B$ & \footnote{Batch, shape, and
crystal structure {\it independent}} &  &  & &  \\ 
$\sigma_D$ & 0.0064$D$ & 0.018$D$ & \footnote{Batch dependent} & X & & & \\
$g$ & 2.00 & 1.94 & $^a$ & & & & \\
$\sigma_g$ & --- & 0.002$g$ & $^b$ & X & & & \\
$J$ & $-7$ gauss & --- & $^b$ & X & & & \\
$\Delta$ & $-20$ gauss & --- & 
\footnote{Shape, size, and crystal structure dependent} &  & X & X & X \\
$\Gamma$ & 86 gauss$^2$ & 203 gauss$^2$ & $^c$ &  & X & X & X \\
$\Lambda$ & $-$156 gauss$^2$ & --- & $^c$ &  & X & X & X \\
\end{tabular}
\end{ruledtabular}
\end{table}

\clearpage

\end{document}